\documentclass[runningheads]{llncs}
\usepackage{graphicx}
\usepackage[export]{adjustbox}
\usepackage{hyperref}
\usepackage{subcaption}
\usepackage{textcomp}

\begin{document}

\newcommand{\so}{schema.org} 

\title{Google Dataset Search by the Numbers}


\author{Omar Benjelloun \and Shiyu Chen \and Natasha Noy}

\authorrunning{O. Benjelloun et al.}

\institute{Google Research, Google, USA \\
\email{\{benjello, shiyuc, noy\}@google.com}}

\maketitle              

\vspace{-0.5cm}
\begin{abstract}
  Scientists, governments, and companies increasingly publish datasets on the
  Web. Google's Dataset Search extracts dataset metadata---expressed using
  schema.org and similar vocabularies---from Web pages in order to make datasets
  discoverable. Since we started the work on Dataset Search in 2016, the number of
  datasets described in schema.org has grown from about 500K to almost
  30M. Thus, this corpus has become a valuable snapshot of data on the Web.
  To the best of our knowledge, this corpus is
  the largest and most diverse of its kind. We analyze this corpus and discuss where the datasets
  originate from, what topics they cover, which form they take, and what people
  searching for datasets are interested in. Based on this analysis, we identify
   gaps and possible future work to help make data more
  discoverable.
\end{abstract}

\section{Dataset Search as a snapshot of datasets on the Web}
\label{sec:introduction}

We live in a data-driven world. Scientists, governments, journalists, commercial
companies, and many others publish millions of datasets online. There are 
thousands of Web sites that publish datasets---some publish a handful, some
publish hundreds of thousands~\cite{chapman.survey.2020}. Google's Dataset
Search\footnote{\url{http://datasetsearch.research.google.com}} is a search
engine for datasets on the Web~\cite{www2019}. It relies on \so\
and similar open
standards to extract the semantics of  dataset metadata and to make it searchable.

Arguably, the mere existence of Dataset Search and its reliance on semantic
markup provided a strong incentive for dataset providers to add such markup to
their  Web pages. Indeed, we have seen an explosive growth of dataset
metadata on the Web since we started the work on Dataset Search. 
In the Fall of 2016,
there were about 500K Web pages that included \texttt{schema.org/Dataset}
markup, with half of them coming from \texttt{data.gov}, the US
Open Government portal~\cite{hendler2012}. Today, there are tens of millions of
such pages, from thousands of sites.

A recent comprehensive survey highlights a variety of approaches to help users
find datasets~\cite{chapman.survey.2020}: these approaches range from searching within a
collection of tables with different schemas~\cite{miller}, to finding data in
repositories, such as Figshare, Zenodo, or DataDryad, to using metadata search
engines, like Dataset Search. There are a number of well respected directories
of dataset publishers (e.g., DataCite~\cite{datacite}, re3data~\cite{re3data},
Scientific Data in Nature~\cite{sdata}), but they inevitably miss new datasets
or repositories~\cite{fenner_2017}. To the best of our knowledge, Dataset Search
is the only collection of dataset metadata that includes all semantically
annotated datasets on the Web.

We chose to rely primarily on \so\ for describing dataset metadata because both
search engines and open-source tools have used it successfully to build an open
ecosystem for various types of content~\cite{guha2016schema}. In recent years,
the scientific community has also embraced it for publishing data, by creating
mappings from other metadata standards to \so. For example, Sansone and
colleagues define a mapping from the DATS standard in the biomedical community
to \so~\cite{dats}. Wang and colleagues use \so\ to describe research-graph
data, comprised of researchers, datasets and scholarly articles~\cite{wang2017}.
Efforts such as bioschemas.org~\cite{gray2017bioschemas} extend \so\ to include
domain-specific terminology and relationships.

In this paper, we analyze the Dataset Search corpus of metadata. As of March 2020, the corpus 
contained 28 million datasets from more than 3,700
sites. While limited to the dataset metadata that is available in \so\ or DCAT,
this corpus contains a sizable snapshot of the datasets on the Web. And because many
researchers and scientists rely on search engines to find
datasets~\cite{Gregory2020Lost}, learning from this corpus can inform both the
work to improve search engines for datasets and, more important, highlight the
gaps in representation and coverage for the community at large. Specifically,
in this paper, we make the following contributions:
\begin{itemize}
\item We present methods for analyzing an organically created corpus of metadata
  for 28 million datasets on the Web (Section~\ref{sec:methods}).
\item We identify a set of research questions that such a corpus can help analyze  and 
present results of the analysis of the corpus (Section~\ref{sec:analysis}).
\item We discuss lessons learned from the corpus analysis (Section~\ref{sec:discussion}).
\end{itemize}

\section{Data collection methods}
\label{sec:methods}

In this section, we describe the methods that we used to collect the metadata
and to prepare it for the analysis in Section~\ref{sec:analysis}. In the
remainder of this paper, we abbreviate the \texttt{schema.org} namespace as
\texttt{so\#} and the \texttt{DCAT} namespace as \texttt{dct\#}.

\subsection{From schema.org and DCAT on the Web to the corpus}

We described the details of the Dataset Search architecture
elsewhere~\cite{www2019}. In brief, Dataset Search relies on the Google Web crawl to find pages that contain dataset metadata and to extract the corresponding triples. A post-processing of the Web crawl data parses RDFa, Microdata, and JSON-LD into a common graph data model, broadly equivalent to W3C's RDF triples~\cite{RDF}. We keep \texttt{so\#Dataset}, \texttt{dct\#Dataset}, and all the related entities and their properties. 
 
We enhance, normalize, and augment this corpus in a variety of ways in order to
provide users with a meaningful search experience. In this section, we focus
only on those processing steps that are relevant to the subsequent data analysis. The processing happens at multiple levels of granularity: At the corpus level, we
ensure that datasets are unique and attempt to remove non-datasets (i.e., pages that include dataset markup, but do not describe datasets). At the
dataset level, we augment the metadata with inferred properties. Finally, at the
property level, we clean up and normalize values.

The \textbf{corpus-level analysis} starts by removing duplicates within each
site~\cite{www2019}. We found that many dataset repositories add
markup both to the dataset landing pages and to the pages that list search results
within that repository. We keep only the former in the corpus through simple
heuristics: When the same dataset (according to values of key properties)
appears on multiple pages, we keep the page that contains only one dataset.
We also remove dataset metadata that does not have values for basic
properties such as title and description~\cite{devSite}. 

At the \textbf{dataset level}, we process the values for properties such as
title and description as well as the terms on the Web page itself in order to
identify the main topics covered by the dataset. We use the topics from
\texttt{re3data.org}~\cite{re3data} and a similar set of topics from the Google
Knowledge Graph~\cite{kg.cacm} as our vocabulary.

In addition, our \textbf{page-level} analysis collects information from the Web page that the dataset originated from, such as the domain of the page and its language.

For \textbf{individual properties}, we normalize, clean, and reconcile values for:
\begin{itemize}
\item \textbf{Data downloads and formats:} We identify the patterns that data
  providers use to represent download information and normalize them to a
  single representation~\cite{www2019}. Providers may specify file formats
  through the \texttt{so\#fileFormat} or \texttt{so\#encodingType} properties.
  When both of these properties are missing, we extract a file
  extension from the data-download URL.
  
\item \textbf{DOIs and compact identifiers:} Persistent citable identifiers,
  such as Digital Object Identifiers (DOIs) and Compact
  Identifiers~\cite{compact.identifiers}, may appear in several properties, such
  as \texttt{so\#identifier}, \texttt{so\#url}, or even \texttt{so\#sameAs}, and
  \texttt{so\#alternateName}. We use regular expressions to find patterns that
  correspond to these identifiers, and look for known prefixes from
  \texttt{identifiers.org} in all of these properties.
  
\item \textbf{Level of access}: Two properties define the level of access of a
  dataset: \texttt{so\#isAccessibleForFree} is a boolean value that indicates
  whether or not the dataset requires a payment.
  \texttt{so\#license} links to a license or specifies one inline. We
  normalize the license information into known classes of licenses, such as
  Creative Commons and open government licenses. Any license that allows redistribution 
  essentially makes the dataset available for free.  We count datasets
  with these licenses as well as datasets with \texttt{so\#isAccessibleForFree}
  set to true as the datasets that are ``open.''
\item \textbf{Providers}: There is some ambiguity in \so\ on how to specify the
  the source of a dataset. We use the
  \texttt{so\#publisher} and \texttt{so\#provider} properties to identify the
  organization that provided the dataset. As with other properties, the value
  may be a string or an \texttt{Organization} object. Wherever possible, we
  reconcile the organization to the corresponding entity in the Google Knowledge
  Graph.
\item \textbf{Updated date:} There are several date properties associated with a
  dataset: \texttt{so\#dateCreated}, \texttt{so\#datePublished},
  \texttt{so\#dateModified} (and similar properties in DCAT). There is little consistency in how dataset publishers distinguish between
  them. However, the most recent value across these dates is usually a reliable
  approximation on when a dataset was last updated. We use several parsers to
  understand dates expressed in common formats and to normalize them. If there is no date in the metadata, we use the date when the Web page itself was last
  updated as a proxy.
\end{itemize}

Finally, to analyze the usage of datasets in the Dataset Search application, we look at logs for two weeks in May 2020. We extract the identifiers of the datasets that appeared in search results, and join them with their metadata to analyze search behavior in aggregate.

All the data and analyses in this paper are based on a snapshot of the Dataset
Search corpus from March 26, 2020. We also compare the status of the corpus with
a version from ten months prior, in June 2019.

\subsection{Limitations of the analysis}
\label{sec:limitations}

While we believe that our corpus is a reasonably representative snapshot of
the datasets published on the Web, we recognize that it has limitations.
Indeed, we have no way of measuring how well the corpus covers all the datasets available on the Web.

First, the corpus  contains only the datasets that have semantic
descriptions of their metadata in \so\ or DCAT. If a dataset page does not have
metadata in a machine-readable format and in a vocabulary that we recognize, it
will not be in our corpus (and will not be discoverable in Dataset Search).

Second, if a dataset page is not accessible to the Google crawler or is not
being crawled for some reason (e.g., because of \texttt{robots.txt}
restrictions), it will not be in our corpus. When the crawler processes
the page, it often needs to execute JavaScript to get the metadata. 
If a page is slow to render, we may not obtain dataset metadata from it.

Third, our methods for inferring new values, such as dataset topics, may be
imprecise, and we have not formally evaluated their accuracy yet. 

Fourth, in our analysis, we ``trust'' what the metadata says. For instance, if a
dataset's metadata says that the dataset is accessible for free, we count it
among the open datasets. In some cases, the reality may be different
when users try to download the dataset.

Finally, a significant amount of pages on the Web are designed for Search Engine Optimization or are simply spam. A page may
have \texttt{so\#Dataset} on it but not actually describe any dataset metadata.
While we do our best to weed out such pages, our techniques are not perfect, and
we cannot be certain that all the datasets in the corpus that we describe are indeed datasets.

\section{Results and Observations}
\label{sec:analysis}

We start with the results of a corpus-level analysis
 (Section~\ref{sec:corpus-level}),
 then look at specific metadata properties
(Section~\ref{sec:inside-the-metadata}) and finally present our observations on the usage of datasets in Dataset Search
(Section~\ref{sec:users}).

\subsection{Corpus-level Analysis}
\label{sec:corpus-level}

Looking at the corpus as a whole, as well as characteristics of the Web pages
that we extracted metadata from, enables us to answer the following
questions.\footnote{Here and elsewhere, ``domain'' refers to ``internet
  domain.''}
\begin{description}
\item [Datasets and domains:] How many datasets does each domain have?
  What is the distribution of the number of datasets by domain? What are the
  domains contributing the largest number of datasets? What are the most common
  top-level domains with datasets, and what fraction of datasets do they
  contribute? We know that many of the datasets come from open-government
  initiatives across the world. But just how many?
\item [Dynamics of the corpus:] How has the corpus of dataset metadata
  grown over time? Pages with datasets inevitably go offline, get moved,
  change their URLs. Can we trace this churn in our corpus? How prevalent is it?
  These numbers give a sense of how metadata on the Web is changing.
\item [Metadata on metadata:] Which fraction of  datasets 
  use schema.org vs DCAT? Which metadata fields are frequently populated, and
  which ones rarely have any values? These numbers give us probably the most
  actionable items in terms of improving metadata quality.
\end{description}

\begin{figure}[tb]
  \begin{subfigure}[T]{.6\linewidth}
    \includegraphics[width=\textwidth]{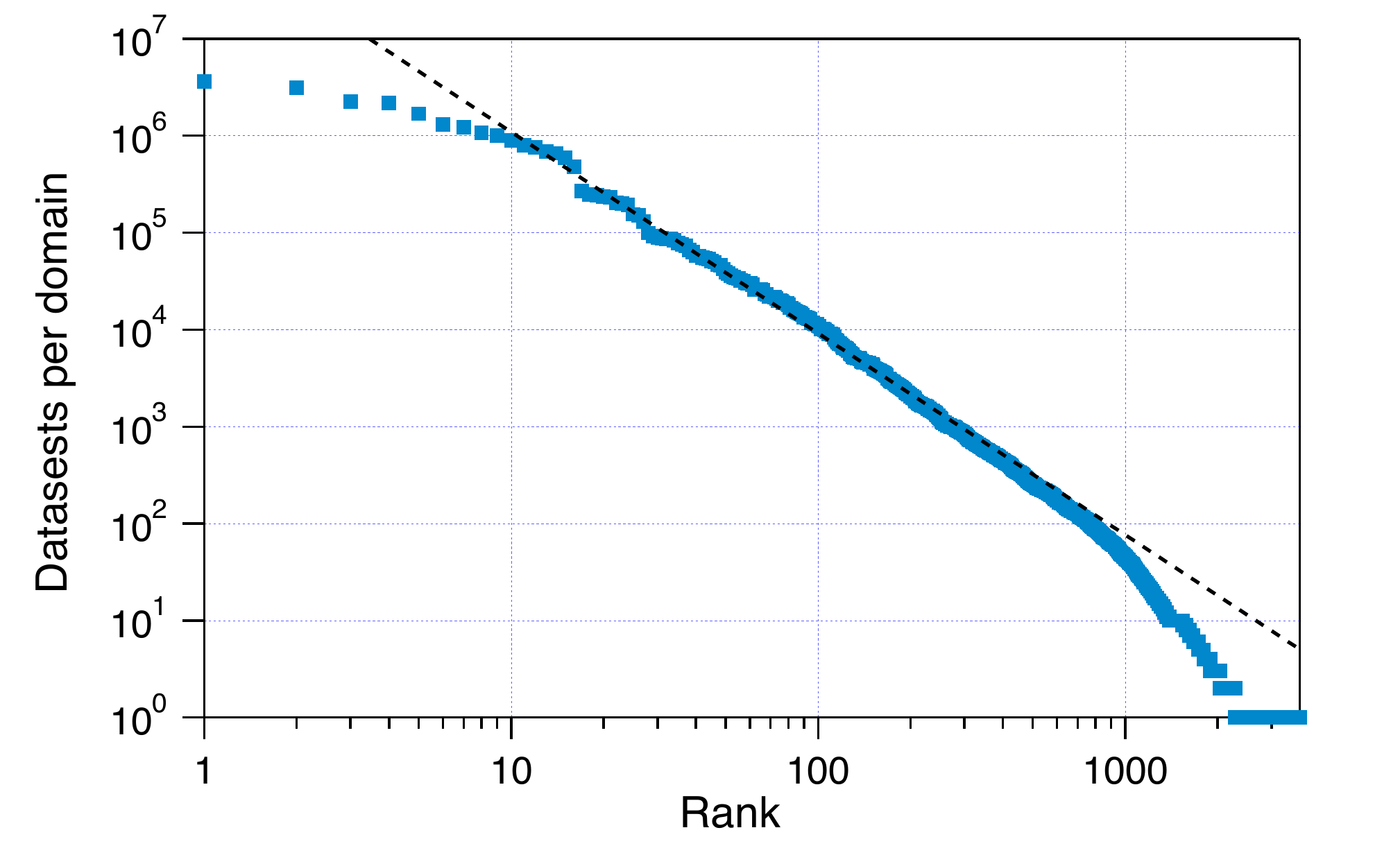}
    \caption[]{}
  \end{subfigure}
\begin{subfigure}[T]{.39\linewidth}
\footnotesize
\begin{tabular}{l r}
  \hline
  \textbf{Domain} & \textbf{Datasets}\\
  \hline
  \href{http://ceicdata.com}{{ceicdata.com}} & 3.7M\\
  \href{http://data.gov}{{data.gov}} & 3.1M\\
  \href{http://hikersbay.com}{{hikersbay.com}} & 2.3M\\
  \href{http://tradingeconomics.com}{{tradingeconomics.com}} & 2.2M\\
  \href{http://knoema.com}{{knoema.com}} & 1.7M\\
  \href{http://figshare.com}{{figshare.com}} & 1.3M\\
  \href{http://stlouisfed.org}{{stlouisfed.org}} & 1.2M\\
  \href{http://datacite.org}{{datacite.org}} & 1.1M\\
  \href{http://thermofisher.com}{{thermofisher.com}} & 1.0M\\
  \href{http://statista.com}{{statista.com}} & 0.9M\\
  \hline
 \end{tabular}
  \caption[]{}
\end{subfigure}
\caption{Number of datasets per domain: (a) A log-log plot of domains and their relative sizes in terms
of the number of datasets. The dotted line shows a fit to power-law  for most of the range
with the coefficient of 2.08 $\pm$ 0.01, which is very close to quadratic fit. 
(b) Domains with the largest number of datasets. These ten domains are responsible for 65\% of the datasets in the corpus.}
\label{fig:domain-size-top}
\end{figure}

\subsubsection{Datasets and domains}
The snapshot that we analyze in the rest of this section, taken on March 26, 2020,
contains 28M datasets from 3,700 domains. The number of datasets per
domain mostly follows a power law distribution, as the logarithmic scale plot
in Figure~\ref{fig:domain-size-top}a shows: A small number of domains publish
millions or hundreds of thousands of datasets, while the long tail of domains
hosts just a handful of datasets. The two domains with
the largest number of datasets (\texttt{ceicdata.com} and \texttt{data.gov}) have
more than 3 million datasets each.  The  ten largest domains (Figure~\ref{fig:domain-size-top}b) account for 65\% of all datasets.

While ``typical'' Web pages about datasets correspond to a single dataset, some
pages may have multiple datasets on them. For instance, a page may describe a
large dataset and break down its components as multiple datasets; or a page may
be dynamically generated in response to a search in a dataset repository. In our
corpus, we found that over 90\% of datasets come from pages that contain exactly
one dataset. Still, more than 1.6M datasets come from pages with more than ten
datasets.

\begin{table}[bt]
\centering
\caption{Number of datasets for the top twenty top-level Internet domains.}
\vspace{0.2cm}
\label{tab:tlds}
\begin{subtable}[]{.4\linewidth}
\begin{tabular}{l r}
\hline
  \vtop{\hbox{\strut \textbf{{Top-level}}}\hbox{\strut \textbf{domain}}} & \quad \vtop{\hbox{\strut \textbf{Number}}\hbox{\strut \textbf{of datasets}}}\\
\hline
\textbf{.com} & 14,956K \\
\textbf{.org} & 4,696K \\
.gov & 3.386K \\
.at & 819K \\
\textbf{.net} & 760K \\
.es & 524K \\
.de & 366K \\
.edu & 293K \\
  .fr & 281K \\
.eu & 263K \\
\hline
\end{tabular}
\end{subtable}
\begin{subtable}[]{.4\linewidth}
\begin{tabular}{l r}
\hline
  \vtop{\hbox{\strut \textbf{{Top-level}}}\hbox{\strut \textbf{domain}}} & \quad \vtop{\hbox{\strut \textbf{Number}}\hbox{\strut \textbf{of datasets}}}\\
\hline
 .ru & 243K \\
 .co & 218K \\
 .nl & 181K \\
 .au & 160K \\
 .pl & 152K \\
 .uk & 144K \\
 .ca & 139K \\
 .io & 79K \\
 .world & 56K \\
 .info & 54K \\
\hline
\end{tabular}
\end{subtable}
\end{table}

Table~\ref{tab:tlds} shows the distribution of datasets by top-level internet
domains. The vast majority of the datasets come from \texttt{.com} domains, but both \texttt{.org}
 and government domains are well represented. For the country-specific
domains, Austria, Spain, Germany, and France are at the top of the list. If we
combine all government domains across the world (\texttt{.gov}, 
\texttt{.gouv.*} , \texttt{.gv.*}, \texttt{.gov.*}, \texttt{.gob.*},
etc.), we find 3.7M datasets on these government domains.

To get a more complete picture of the international coverage of datasets,
Table~\ref{tab:languages} breaks them down by language, as specified by or
extracted from the Web pages that contain them. More than 18M
datasets, or 64\% are in English, followed by datasets in Chinese and
Spanish, both of which are growing faster than datasets in English. Note that these
numbers do not capture the nuance of specific schema.org property values using
multiple languages.

\begin{table}[tbh]
\centering
\caption{Number of datasets by language and the \% change between  June 2019 and March
  2020.}
\vspace{0.2cm}
\label{tab:languages}
\begin{tabular}[]{l r r}
  \hline
  \textbf{Language} & \quad \vtop{\hbox{\strut \textbf{Number}}\hbox{\strut \textbf{of datasets}}} & \quad \textbf{\% increase} \\
\hline
English & 18,650K & 67\% \\
Chinese & 1,851K & 82\% \\
Spanish & 1,485K & 70\% \\
German & 743K & 74\% \\
French & 492K & 76\% \\
Arabic & 435K & 75\% \\
Japanese & 404K & 72\% \\
Russian & 354K & 65\% \\
Portuguese & 304K & 69\% \\
Hindi & 288K & 70\% \\
\hline
\end{tabular}
\end{table}

\subsubsection{Dynamics of the corpus}

\begin{figure}[bth]
\begin{subfigure}{.65\textwidth}
  \centering
  \includegraphics[width=0.9\textwidth]{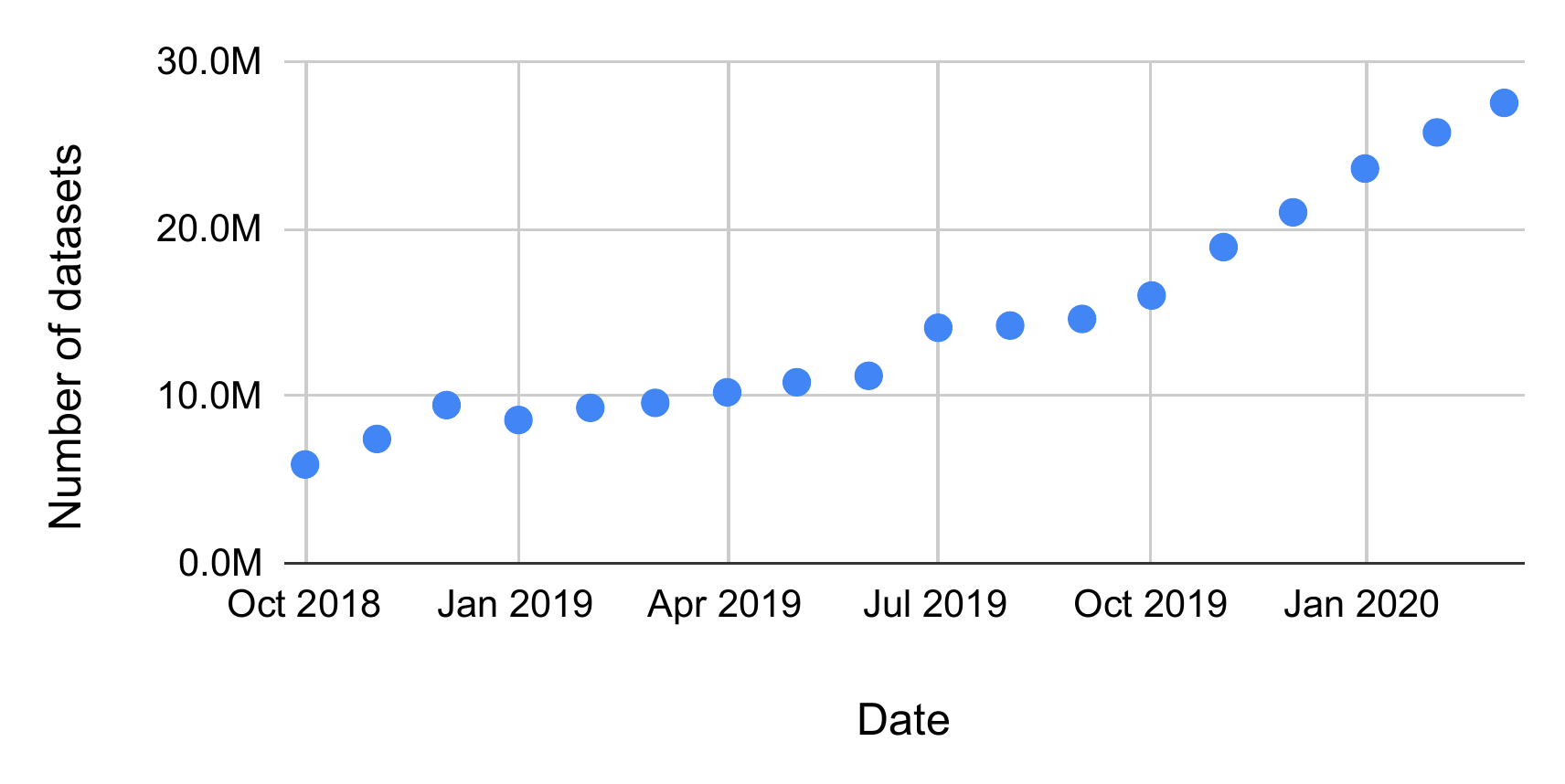}
   \caption[]{}
 \end{subfigure}
\begin{subfigure}{.3\textwidth}
  \centering
  \includegraphics[width=\textwidth]{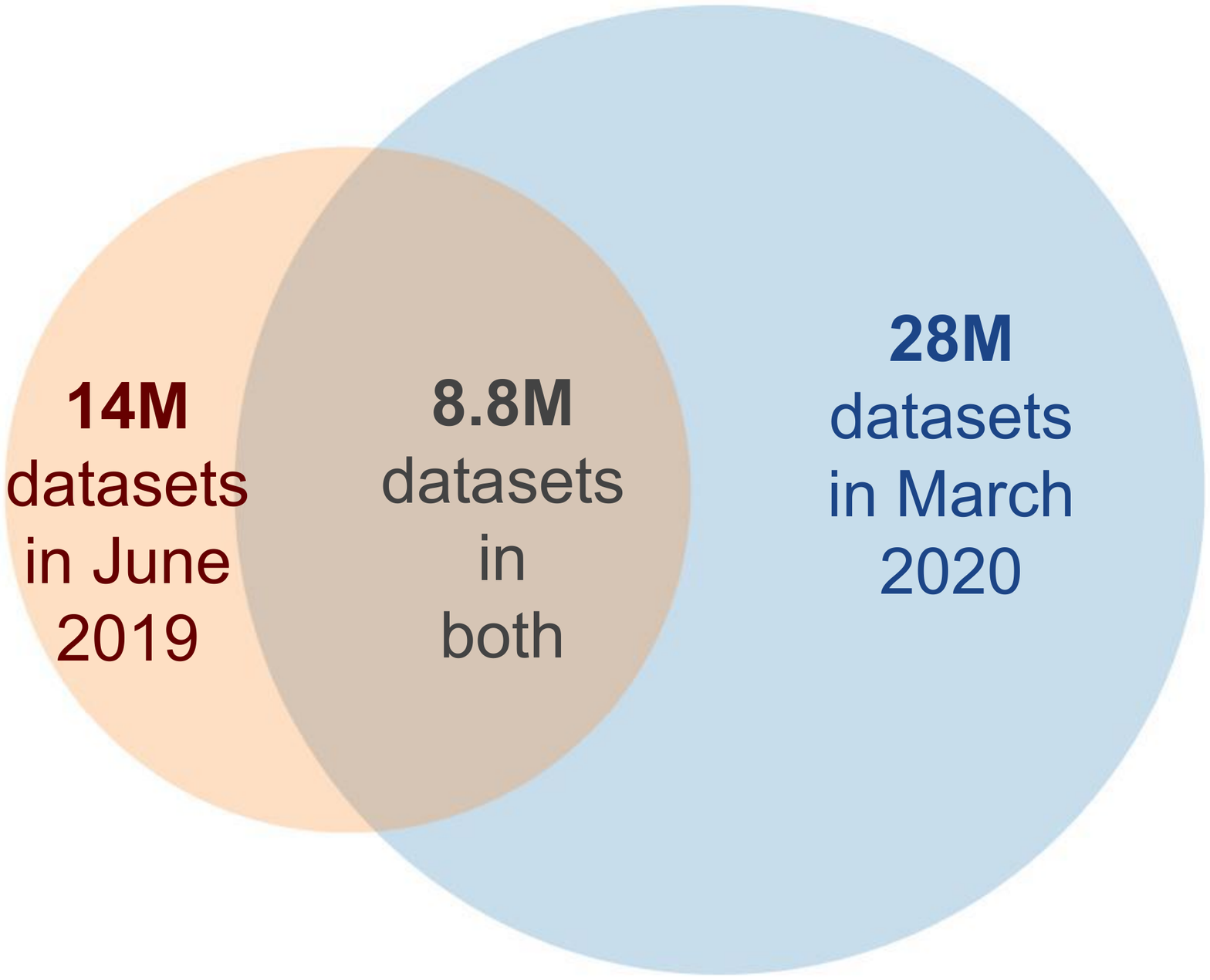}
   \caption[]{}
\end{subfigure}
\caption{Corpus dynamics: (a)
  Growth of datasets since the beta launch in September 2018. The number of datasets grew from 
  6M to 28M (b) Changed
  URLs in  the corpus between June 2019 and March 2020: 8.8M URLs of the 14M in June 2019 remained in the corpus. The rest have either disappeared or changed.}
\label{fig:growth}
\end{figure}

The next question that we study is the change in the corpus over time.
Figure~\ref{fig:growth}a shows the growth in the number of datasets since the beta launch of Dataset Search, in September 2018. The number of datasets has grown steadily, from about 6M then to 28M in March, 2020. We have reported earlier~\cite{www2019} that,
day-to-day, about 3\% of the datasets are deleted from our index while 7-10\%
new datasets are added. Enterprise data repositories have a similarly large
level of churn~\cite{goods}. Figure~\ref{fig:growth}b shows the results of comparing
the URLs between snapshots from June
2019 and March 2020, when the corpus almost doubled in size: Only 8.8M
of the 14M URLs in the June 2019 corpus, or 63\%, are still there in March 2020. 
The other 5.4M are no longer at the same location---or may no longer be in the 
corpus at all. This dynamic indicates a very high level of churn.

\begin{figure}[bt]
\begin{subfigure}{.5\textwidth}
  \centering
  \includegraphics[width=\textwidth]{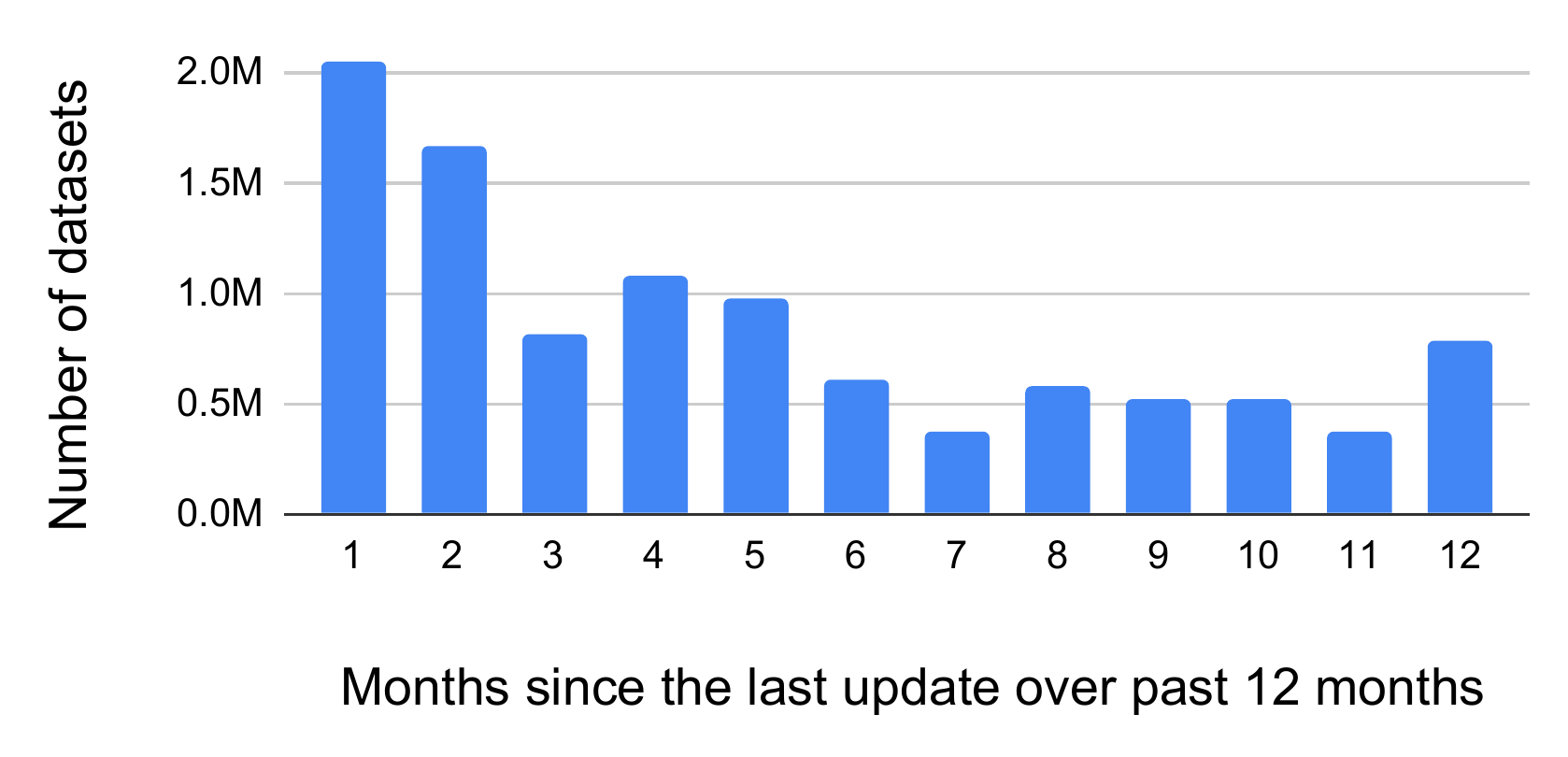}
   \caption[]{}
\end{subfigure}
\begin{subfigure}{.5\textwidth}
  \centering
  \includegraphics[width=\textwidth]{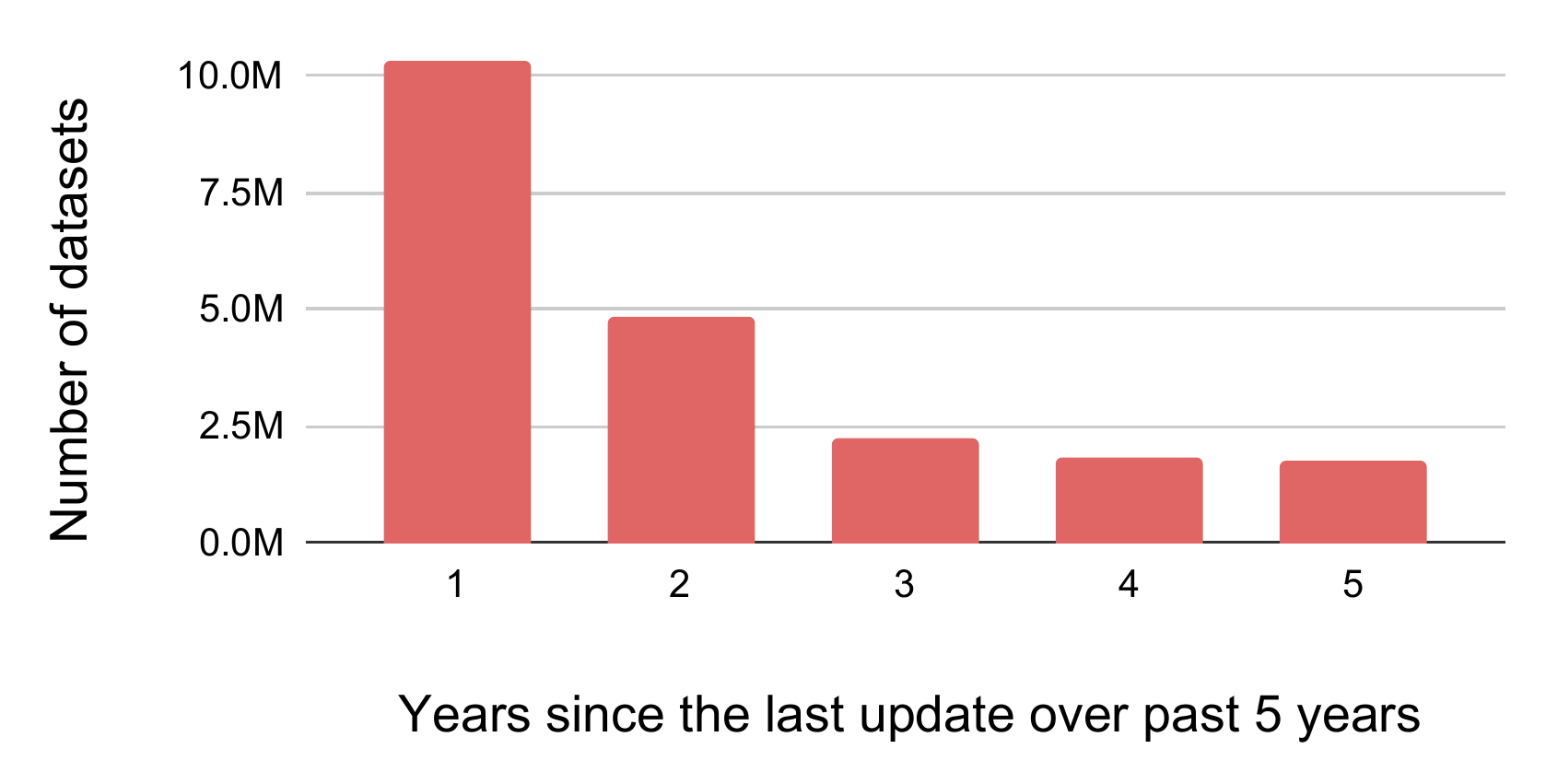}
   \caption[]{}
\end{subfigure}
\caption{Distribution of the date when a dataset was last updated: (a) at monthly
  granularity over the past year; (b) at yearly granularity over the last five
  years. Note that we have this information only for 85\% of datasets.}
\label{fig:updates}
\end{figure}

Figure~\ref{fig:updates} captures the updates to individual datasets. 
14M datasets, or 50\%, have a value for at least one of the date properties in metadata 
 (Section~\ref{sec:methods}). For an additional 10.7M datasets, we were able
to determine the date when the Web page was modified. Out of these 24.7M datasets with a known date of last update, 21M datasets, or 85\%, have this date within the last 5 years (Figure~\ref{fig:updates}). 
The short-term distribution (Figure~\ref{fig:updates}a) shows that more datasets were last updated within the last  month  than in any other month in the past year. Looking at the long-term distribution (Figure~\ref{fig:updates}b),
49\% of datasets were last updated within the past year.

\subsubsection{Metadata on metadata}

While \so\ is our primary semantic vocabulary for dataset description, we
also understand and map basic DCAT properties. However, we found that fewer
than 1\% of datasets use the DCAT vocabulary.

Table~\ref{tab:fields} shows which fraction of datasets have values for specific
properties. We require datasets to have a title and
description~\cite{devSite};
hence, their 100\% coverage in our corpus. Because we normalize and reconcile
values from different properties, the properties in Table~\ref{tab:fields} do
not always directly correspond to schema.org or DCAT predicates. For instance,
we combine \texttt{so\#publisher} and \texttt{so\#creator} into ``provider''
because we observed that data owners do not really distinguish between the
different semantics.

\begin{table}[htb]
\centering
\caption{Percentage of datasets with specific properties. Column 2 lists the source
predicates for each property. Properties not listed in the table have values in fewer
than 1\% of the datasets.}
\vspace{0.2cm}
\label{tab:fields}
\begin{tabular}{l l r}
  \hline
  \textbf{Property} & \textbf{Source predicates} & \textbf{Percentage} \\
  \hline
  description & \texttt{so\#description}, \texttt{purl\#description} & 100.00\% \\
  title & \texttt{so\#name}, \texttt{purl\#title} & 100.00\% \\
  provider & \texttt{so\#publisher}, \texttt{so\#provider}, \texttt{purl\#publisher} & 84.59\% \\
  keywords & \texttt{so\#keywords}, \texttt{dct\#keyword}, \texttt{purl\#keyword}  & 80.08\% \\
  URL & \texttt{so\#url}, \texttt{dct\#accessurl}, \texttt{dct\#landigpage}  & 68.30\% \\
  temporal coverage & \texttt{so\#temporalCoverage}, \texttt{so\#temporal}, \texttt{purl\#temporal}  & 45.41\% \\
  data download & \texttt{so\#distribution}, \texttt{dct\#distribution} & 44.34\% \\
  spatial coverage & \texttt{so\#spatialCoverage}, \texttt{so\#spatial},  \texttt{purl\#spatial}  & 38.69\% \\
  date modified & \texttt{so\#dateModified}, \texttt{purl\#modified} & 37.46\% \\
  license & \texttt{so\#license} and \texttt{so\#license} on \texttt{so\#distribution} & 34.80\% \\
  date published & \texttt{so\#datePublished},  \texttt{purl\#published}  & 30.83\% \\
  catalog & \texttt{so\#includedInCatalog} & 29.74\% \\
  variable & \texttt{so\#variableMeasured}, \texttt{dct\#theme} & 20.90\% \\
  authors & \texttt{so\#author}, \texttt{so\#creator}  & 14.12\% \\
  same\_as & \texttt{so\#sameAs}, \texttt{rdf\#same\_as} & 12.72\% \\
  date created & \texttt{so\#dateCreated} & 9.62\% \\
  alternate name & \texttt{so\#alternateName}, \texttt{rdf-schema\#label} & 3.40\% \\
  is accessible for free & \texttt{so\#isAccessibleForFree}  & 3.04\% \\
  \hline
\end{tabular}
\end{table}

\subsection{Inside the metadata}
\label{sec:inside-the-metadata}

We focus on the  properties that we found most
informative in understanding the corpus---or that we got asked about most often
(e.g., what are the formats for the available datasets). We do not try to
provide an exhaustive description of the ranges of values for every property.

\subsubsection{Topics}

\begin{figure}[bh!]
\includegraphics[width=\textwidth]{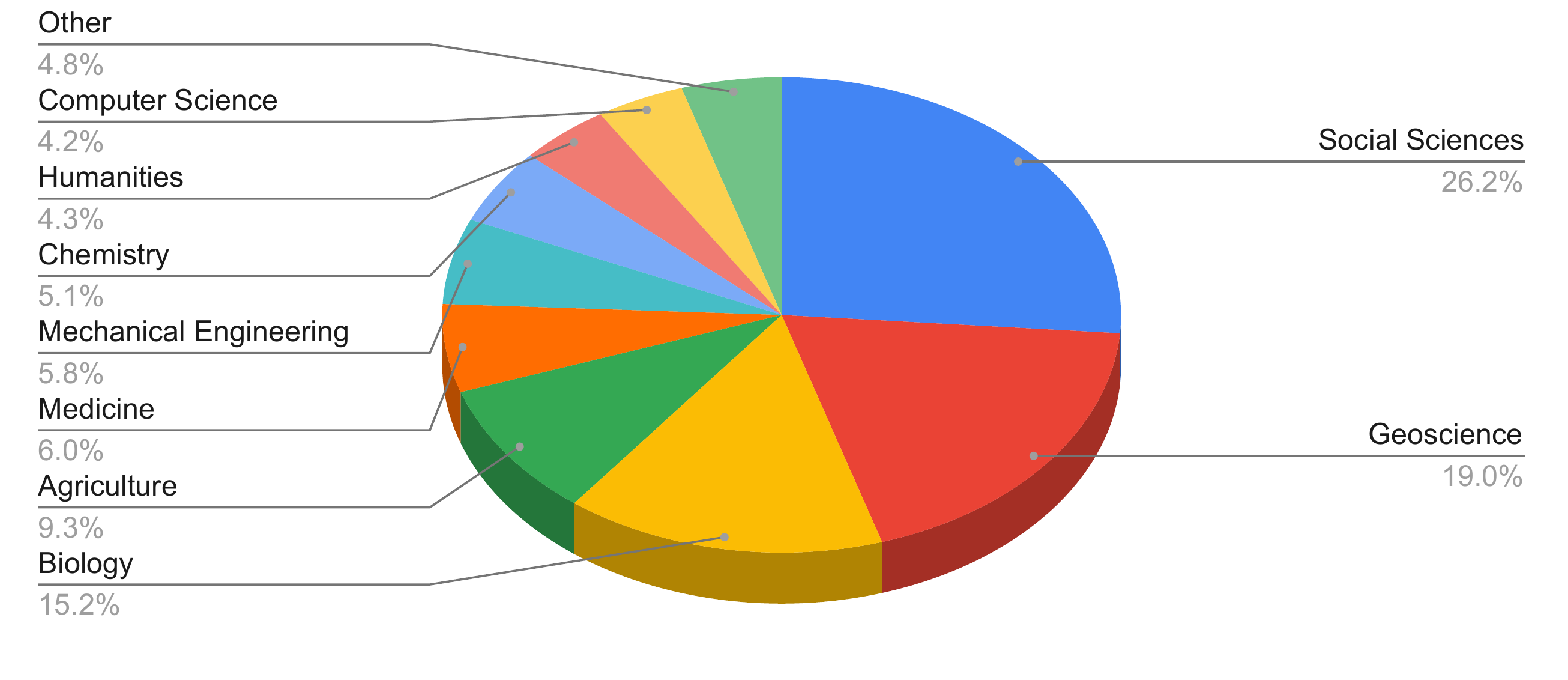}
\vspace{-0.5cm}
\caption{Distribution of datasets by broad coverage topic, inferred from dataset metadata and 
the Web page itself.}
\label{fig:topics}
\end{figure}

Figure~\ref{fig:topics}
shows the distribution of the topics that datasets are associated with. We generate the
topics automatically, based on dataset titles and descriptions as well as the
text on the page where the dataset came from. The two largest topics are
geosciences and social sciences---the areas that  we focused on specifically before
the beta launch in September 2018.

\subsubsection{Data downloads}

Most users who search for datasets ultimately want the data itself and not just
its metadata. Datasets can specify means to download their data via the
\texttt{so\#distribution} property. 

\paragraph{Availability:}
\label{sec:availability}

Only 44\% of datasets specify a data download link in their metadata
(Table~\ref{tab:fields}). Looking at the origin of datasets with downloads, we
found that 85\% of them are provided by just 10 domains.


\paragraph{Data types and formats:}
\label{sec:data-types-and-formats}

Zooming in on the subset of datasets that specify a data download, what are the
broad categories of content and their relative prevalence? To answer this
question, we first extract the file format of data downloads, and then bucket
them into categories. For bucketing, we defined a high-level classification
inspired by Elsevier DataSearch,\footnote{\url{datasearch.elsevier.com/}} and
created a mapping from the file formats found in the data to the target
categories (Table~\ref{tab:content-types}).

\begin{table}[tb]
  \centering
  \caption{Number of datasets by content type. The counts are based on \texttt{so\#fileFormat} or \texttt{so\#encodingType} properties and file extensions. Note that some datasets have multiple distribution formats. Therefore, the total number of entries here is larger than the 12M datasets with data downloads.}
  \vspace{0.2cm}
  \label{tab:content-types}
  \begin{tabular}{l r r l}
    \hline
    \textbf{Category} & \vtop{\hbox{\strut \textbf{Number}}\hbox{\strut \textbf{of datasets}}} & \quad \vtop{\hbox{\strut \textbf{\% of}}\hbox{\strut \textbf{total}}} & \quad \textbf{Sample formats}\\
    \hline
    Tables & 7,822K & 37\% & \quad CSV, XLS \\
    Structured & 6,312K & 30\% & \quad JSON, XML, OWL, RDF \\
    Documents & 2,277K & 11\% & \quad PDF, DOC, HTML \\
    Images & 1,027K & 5\% & \quad JPEG, PNG, TIFF \\
    Archives & 659K & 3\% & \quad ZIP, TAR, RAR \\
    Text & 623K & 3\% & \quad TXT, ASCII \\
    Geospatial & 376K & 2\% & \quad SHP, GEOJSON, KML \\
    Computational biology & 110K & $<$1\% & \quad SBML, BIOPAX2, SBGN \\
    Audio & 27K &  $<$1\% & \quad WAV, MP3, OGG \\
    Video & 9K &  $<$1\% & \quad AVI, MPG \\
    Presentations & 7K &  $<$1\% & \quad PPTX \\
    Medical imaging & 4K &  $<$1\% & \quad NII, DCM \\
    Other categories & 2,245K & 11\% &  \\
    \hline
  \end{tabular} 
\end{table} 

Tables in CSV or XLS formats are the most common type of data (37\%), followed by
structured formats such as JSON and XML (30\%) and documents in PDF or DOC
format (11\%). The latter category is problematic for many applications, as it
is not machine readable. Audio, video, and medical imaging formats all
constitute less than 1\% of the datasets.

A subset of datasets that is of interest to the Semantic Web community is the
datasets that contain graph data. We can approximate this number by summing
over common formats: owl, rdf, xml+rdf, sparql, and so on. Together, they represent
only 0.54\% of datasets with downloads. Semantic web data is largely
under-represented among datasets that use Semantic Web methods to describe metadata.





\subsubsection{Making data citable}

As we hope that datasets themselves become first-class citizens of the
scientific discourse, we must develop mechanisms to reference and cite them.
Scientists commonly use digital object identifiers (DOIs) and compact
identifiers (provided by services such as identifiers.org) for this purpose. We
extract DOIs and compact identifiers from the dataset URLs or the values of the
\texttt{so\#url} property, as well as from \texttt{so\#sameAs} and
\texttt{so\#identifier} properties. About 11\% of the datasets in the corpus (or
\url{~}3M) have DOIs; about 2.3M of those come from two sites,
datacite.org and figshare.com (Table~\ref{tab:dois}a). Only a tiny fraction,
0.45\% of the datasets, have compact identifiers (Table~\ref{tab:dois}b).

\begin{table}[tb]
  \centering
  \caption{Datasets with DOIs and compact identifiers}
  \label{tab:dois}
  \begin{subtable}[t]{.45\linewidth}
    \caption{Top ten providers of datasets with DOIs.}
    \begin{tabular}{l r}
    \hline
    \textbf{Domain} & \vtop{\hbox{\strut \textbf{Datasets}}\hbox{\strut \textbf{with DOIs}}}\\
    \hline
    \href{http://figshare.com}{{figshare.com}} & 1,300,745 \\
    \href{http://datacite.org}{{datacite.org}} & 1,070,066 \\
    \href{http://narcis.nl}{{narcis.nl}} & 118,210 \\
    openaire.eu & 109,149 \\
    \href{http://datadiscoverystudio.org}{{datadiscoverystudio.org}} &
                                                                       72,063 \\
    \href{http://osti.gov}{{osti.gov}} & 62,923 \\
    \href{http://zenodo.org}{{zenodo.org}} & 49,622 \\
    \href{http://researchgate.net}{{researchgate.net}} &
                                                         41,494 \\
    \href{http://da-ra.de}{{da-ra.de}} & 39,318 \\
    \hline
  \end{tabular}
  \end{subtable}
  \quad
  \begin{subtable}[t]{.45\linewidth}
    \caption{Providers with more than 100 datasets with compact identifiers.}
    \begin{tabular}{l r}
  \hline
  \textbf{Domain} & \vtop{\hbox{\strut \textbf{Datasets with}}\hbox{\strut \textbf{compact identifierss}}} \\
  \hline
  \href{http://neurovault.org}{{neurovault.org}} & 73,869 \\
  \href{http://alliancegenome.org}{{alliancegenome.org}} & 29,204 \\
  \href{http://datacite.org}{{datacite.org}} & 14,982 \\
  \href{http://openaire.eu}{{openaire.eu}} & 4,262 \\
  \href{http://scicrunch.org}{{scicrunch.org}} & 1,522 \\
  \href{http://mcw.edu}{{mcw.edu}} & 517 \\
  \href{http://duke.edu}{{duke.edu}} & 306 \\
  \hline
    \end{tabular}
    \end{subtable}
\end{table}

\subsubsection{Data Providers}

While internet domains provide the conduit that brings datasets to users, the
semantic provenance of datasets is more accurately captured by the notion of
``provider.'' Domains and providers often align, but  they also may differ when a
provider hosts their datasets on a platform different from their own Web site. About
84\% of all datasets specify a provider and there are about 100k  distinct data
providers in the corpus. The top 3 providers are CEICdata.com, Knoema, and the U.S.
Geological Survey. The top 20 providers account for 78\% of the total datasets.
Most of them are the hosts of the top domains in Figure~\ref{fig:domain-size-top}b.
However, 87\% of the providers are ``small'' providers, who publish
fewer than 10 datasets each.

\subsubsection{How open is the data}

Finally, we analyze the licenses and availability of datasets. Dataset Search
does not require the data to be open; only the metadata must be 
accessible to the crawler. Publishers specify access requirements for a dataset 
via the \texttt{so\#license} property, the
\texttt{so\#isAccessibleForFree} boolean property, or both. About a third of the
datasets (34\%) specify license information and 3\% of the datasets have the
value for \texttt{so\#isAccessibleForFree} (Table~\ref{tab:fields}).
Of the datasets that specify a license, we were able to recognize a known
license in 72\% of the cases. Those licenses include Open Government licenses for the US
and Canada, Creative Commons licenses, and several Public Domain licenses. We
found that for 89.5\% of these datasets either the
\texttt{so\#isAccessibleForFree} bit is set to true or their license is a  license
that allows redistribution, or both. In other words, almost 90\% of these datasets
are available for free. And of these open datasets, 5.6M, or 91\%, allow commercial reuse.

\subsection{What do users search for}
\label{sec:users}
\begin{figure}
 \includegraphics[width=\textwidth]{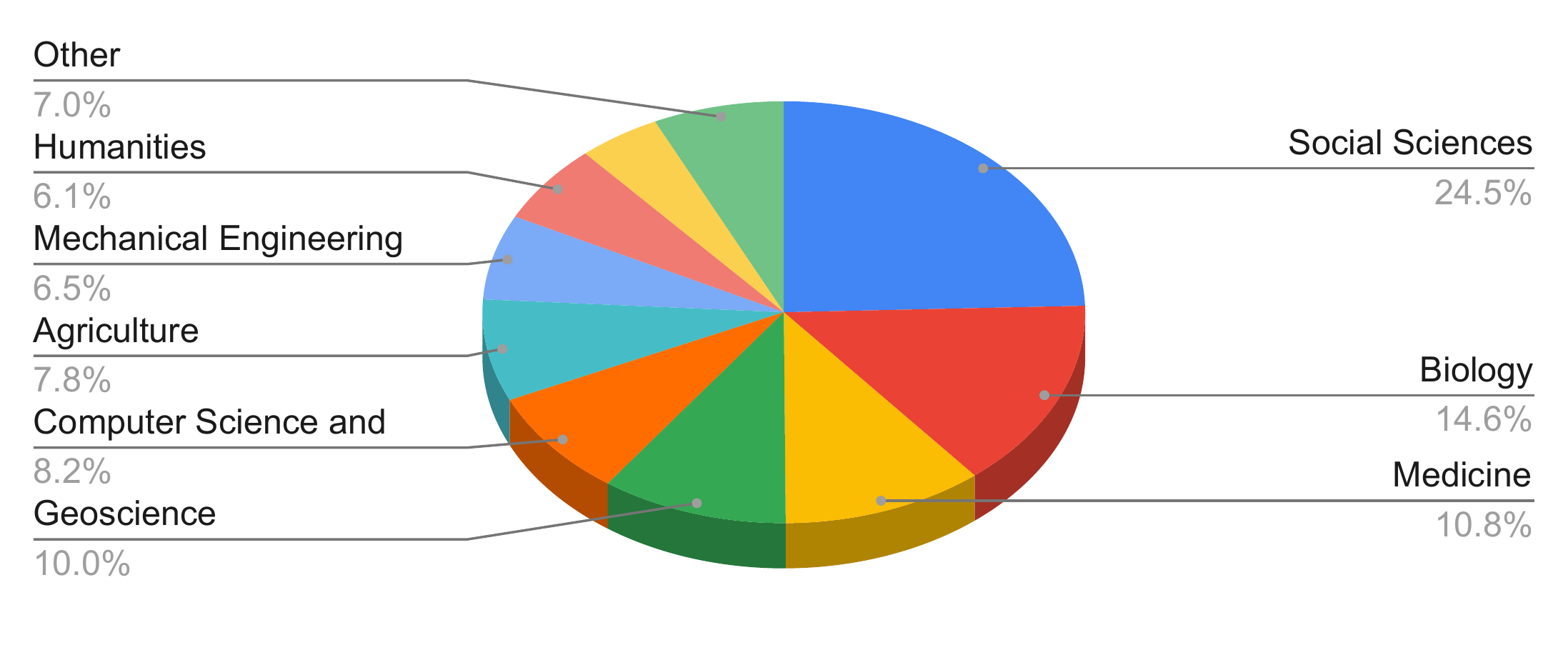}
 \caption{Topic distribution of datasets that appeared in search results over 14
   days in May 2020.}
 \label{fig:topic-frequency}
 \end{figure}

Finally, we look at what 
Dataset Search users search for. Overall, 2.1M
unique datasets from 2.6K domains appeared in the top 100 search results over
 14 days in May 2020. Figure~\ref{fig:topic-frequency} shows the topics for these
datasets. Note that the distribution of topics in
Figure~\ref{fig:topic-frequency} is different from the one for the corpus as a whole
(Figure~\ref{fig:topics}), with geosciences, for instance, taking up a much
smaller fraction; conversely, biology and medicine take a larger fraction
relative to their share of the corpus. We are writing this paper
during the Coronavirus pandemic; this timing likely explains the increased share of the
biology and medicine datasets.

\section{Discussion}
\label{sec:discussion}

We start our discussion by highlighting the results we found surprising or
counter-intuitive. We then focus on the results that point to future work around
improving the quality of metadata in general and provenance information in
particular.

\subsection{What surprised us in the data}
\label{sec:what-surprised-us-in-the-data}

We do not attempt to discuss every table and graph in
Section~\ref{sec:analysis}. Rather, we focus on the results that require some
explanation or discussion.

\textbf{Licenses and access}: Only 34\% of the datasets provide any licensing
information, through the \texttt{so\#license} property for dataset
or distribution. 
When no license is specified, the user technically must assume that
they cannot reuse the data. Thus, adding licensing information, and, ideally,
adding as open a license as possible, will greatly improve the reusability of
the data~\cite{Carbon.licenses}. At the same time, we were encouraged to see
that in the vast majority of datasets that specify a license were available for
free, allowed redistribution under certain conditions, and almost always allowed
reuse for both commercial and non-commercial purposes. While 44\% is still
relatively low, this number is significantly higher than the 9\% of linked-data
datasets with licenses that Schmachtenberg and colleagues found in their 2014
survey~\cite{Schmachtenberg}.

\textbf{Availability of data downloads}: We found that only 44\% of datasets
specify a data download link in their metadata. This number is surprisingly low,
because datasets are merely containers for data. A possible explanation is that
Webmasters (or dataset-hosting platforms) fear that exposing the data-download
link through Schema.org metadata may lead search engines or other applications
to give their users direct access to downloading the data, thus ``stealing''
traffic from their Web site. Another concern may be that data needs the proper
context to be used appropriately (e.g., methodology, footnotes, license
information), and providers feel that only their Web pages can give the complete
picture. In Dataset Search, we made the decision not to show download links as
part of dataset metadata so that users can get the full context from the
publisher's Web site before downloading the data.

\textbf{Quality and completeness of the metadata}: As Table~\ref{tab:fields}
shows, the majority of metadata properties have under 50\% coverage. While
properties such as spatial or temporal coverage  apply only to specific domains,
others such as authors, variables being measured, updated date, licensing and
download information are general, and need to become much more prevalent for
metadata to be truly descriptive. Meusel and colleagues analyzed schema.org
adoption across the Web in 2015~\cite{LOD.survey} and found that data providers
both adopt and drive changes in the schema.

\textbf{Effect of outreach}: We discussed elsewhere~\cite{www2019} that reaching
out to specific communities and finding influencers there was key to
bootstrapping the corpus in the first place. At the time, we focused on
geosciences and social sciences. Since then, we have allowed the corpus to grow
organically. We were surprised to see that, even after the corpus has grown
manifold, the communities that we reached out to early on are still dominating
the corpus: 45.2\% of the datasets are from these two disciplines. Of course, this
dominance may be due to other factors, such as differences in culture across
communities. For instance, geosciences have been particularly successful in
making their data FAIR~\cite{Stall.Nature}.

\textbf{Persistent identifiers and URLs}: Many scientific disciplines have come
to a consensus (or have been compelled by funding agencies and academic
publishers) that it is important to publish data and to cite it~\cite{FAIR}.
There are, for example, peer-reviewed journals dedicated to publishing valuable
datasets, such as Nature Scientific Data~\cite{sdata}, and efforts such as
DataCite~\cite{datacite}, that provide digital object identifiers (DOIs) for
datasets and both encourage and enable scientists to publish their datasets. For
datasets to become first-class citizens in scientific discourse, they must be
citable. Unfortunately, Figure~\ref{fig:growth}b shows that URLs of datasets are
not persistent: 37\% of URLs that had dataset metadata in June 2019 either do
not have the metadata or are no longer accessible in March 2020. This high level
of churn argues strongly for the use of persistent identifiers for datasets,
such as DOIs and compact identifiers. This practice is now widespread for
publications, and we argue that it should become just as widespread for
datasets.

\textbf{(Not) eating our own dogfood}. Fewer than 1\% of datasets in our corpus
are in linked-data formats. The Dataset Search approach relies on semantic-web
technologies such as DCAT and schema.org. At the same time, the Semantic Web
community is either not producing enough data, not sharing it, or not adding
semantic metadata to it. From profiling efforts
(e.g.,~\cite{Ellefi,LOD.survey}), it seems that the problem is the latter: there
is plenty of shared data that researchers produce, but the final step of
describing it appears to be less common.

\subsection{Future work}
\label{sec:future-work}

\subsubsection{Improving metadata quality}

Throughout this analysis, we found many places where metadata was missing,
formatted wrongly, not normalized, and so on. We discuss a few possible
approaches to improve the quality of metadata:
\begin{description}
\item [Automated techniques:] We continue to develop better techniques to
  automatically clean, normalize, and reconcile dataset metadata. These techniques
  are hard to do at scale given the heterogeneity of the data, and they will
  never be perfect. The benefits of these techniques are often
  application-specific, and there are no easy mechanisms to share them back with
  the community.
\item [Feedback to publishers:] We can let the owners of datasets know
  that their dataset metadata can be improved. Google tools such as Search
  Console and the Structured Data Testing tool already highlight some of these
  issues. One could also consider developing interactive tools to create and
  validate dataset metadata, which could be integrated into popular dataset
  management CMSs or hosting platforms.
\item [Crowdsourcing:] Why not let users of datasets fix their metadata or point to
  possible improvements? One option would be to provide that
  functionality in the Dataset Search tool, which would then funnel suggestions
  to the publishers. Vrande{\v{c}}i{\'c}~\cite{vrandecic.akts} proposed using WikiData to
  crowdsource the definition of metadata for popular datasets that the community
  cares about.
\end{description}

\subsubsection{Improving provenance information}

Many datasets are duplicates of other datasets, or derived from other datasets.
Knowing these relationships is important not just for Dataset Search, but to any
user of datasets who cares about data provenance. These relationships are also
critical to giving dataset publishers credit when their data is reused:
derivative datasets are akin to citations of papers. While schema.org provides
properties to describe these relationships, namely \texttt{so\#sameAs} and
\texttt{so\#isBasedOn}, the usage of these properties is low. How can we improve
the coverage of this lineage information?
\begin{description}
\item [Automated detection:] We already detect duplicate datasets and
  cluster them~\cite{www2019}. Identifying that a dataset is derived from
  other datasets is a much more difficult problem, but it may be
  feasible in restricted cases (e.g., specific forms of data, such as tables or
  images, and limited transformation operations).
\item [License requirements:] Most data licenses require citation when
  re-using a dataset, however there is no obligation to make that citation
  machine readable. There would be huge value in requiring the usage of
  schema.org properties when citing datasets.
\item [Community incentives:] Data provenance can help complete the
  picture of the usefulness and impact of datasets, together with paper
  citations, application usage, etc. How can we incentivize broad adoption of
  data provenance in the scientific and open data communities~\cite{herschel2017survey}?
\end{description}

\section{Conclusions}
\label{sec:conclusions}

In this paper, we analyzed the corpus of dataset metadata used in Google's
Dataset Search, a search engine over datasets on the Web. While it has
limitations (Section~\ref{sec:limitations}), it is a large snapshot of datasets
on the Web in a variety of disciplines. Our analysis shows that datasets on the
Web are very diverse, with no one discipline truly dominating; there are
datasets with semantic markup in Web sites from any country and
in any language. We have observed an explosive growth over the last three years.

Yet, metadata still leaves a lot to be desired if data is truly to become a
first-class citizen in scientific discourse: We need tools to ensure that the
metadata is more complete and mechanisms to encourage the use of licensing
information for data and persistent identifiers. And the Semantic Web community
needs to eat its own dogfood by adding semantic metadata to its datasets. 

\bibliographystyle{splncs03}

\begin{thebibliography}{10}
\providecommand{\url}[1]{\texttt{#1}}
\providecommand{\urlprefix}{URL }
\providecommand{\doi}[1]{https://doi.org/#1}

\bibitem{Ellefi}
Ben~Ellefi, M., Bellahsene, Z., Breslin, J.G., Demidova, E., Dietze, S.,
  Szyma{\'n}ski, J., Todorov, K.: {RDF} dataset profiling---a survey of
  features, methods, vocabularies and applications. Semantic Web
  \textbf{9}(5),  677--705 (2018)

\bibitem{Carbon.licenses}
Carbon, S., Champieux, R., McMurry, J.A., Winfree, L., Wyatt, L.R., Haendel,
  M.A.: An analysis and metric of reusable data licensing practices for
  biomedical resources. PLOS ONE  \textbf{14}(3) (2019).
  \doi{10.1371/journal.pone.0213090}

\bibitem{chapman.survey.2020}
Chapman, A., Simperl, E., Koesten, L., Konstantinidis, G., Ib{\'a}{\~n}ez,
  L.D., Kacprzak, E., Groth, P.: Dataset search: a survey. The VLDB Journal
  \textbf{29}(1),  251--272 (2020)

\bibitem{fenner_2017}
Fenner, M., Crosas, M., Grethe, J., et~al.: A data citation roadmap for
  scholarly data repositories. bioRxiv  (2017). \doi{10.1101/097196}

\bibitem{devSite}
Datasets: Search for developers.
  \url{https://developers.google.com/search/docs/data-types/dataset}

\bibitem{gray2017bioschemas}
Gray, A.J., Goble, C.A., Jimenez, R.: Bioschemas: From potato salad to protein
  annotation. In: Intl Semantic Web Conf. (Posters, Demos \& Industry Tracks)
  (2017)

\bibitem{Gregory2020Lost}
Gregory, K., Groth, P., Scharnhorst, A., Wyatt, S.: Lost or found? discovering
  data needed for research. Harvard Data Science Review  (4 2020).
  \doi{10.1162/99608f92.e38165eb}

\bibitem{guha2016schema}
Guha, R.V., Brickley, D., Macbeth, S.: Schema.org: evolution of structured data
  on the web. Communications of the ACM  \textbf{59}(2),  44--51 (2016)

\bibitem{goods}
Halevy, A., Korn, F., Noy, N.F., Olston, C., Polyzotis, N., Roy, S., Whang,
  S.E.: Goods: Organizing {Google's} datasets. In: ACM SIGMOD (2016)

\bibitem{hendler2012}
Hendler, J., Holm, J., Musialek, C., Thomas, G.: {US} {G}overnment {L}inked
  {O}pen {D}ata: Semantic.data.gov. IEEE Intelligent Systems  \textbf{27}(3),
  25–31 (May 2012). \doi{10.1109/MIS.2012.27}

\bibitem{herschel2017survey}
Herschel, M., Diestelk{\"a}mper, R., Lahmar, H.B.: A survey on provenance: What
  for? what form? what from? The VLDB Journal  \textbf{26}(6),  881--906 (2017)

\bibitem{re3data}
Kindling, M., van~de Sandt, S., R{\"{u}}cknagel, J., Schirmbacher, P., Pampel,
  H., Vierkant, P., Bertelmann, R., Kloska, G., Scholze, F., Witt, M.: The
  landscape of research data repositories in 2015: {A} re3data analysis. D-Lib
  Magazine  \textbf{23}(3/4) (2017). \doi{10.1045/march2017-kindling}

\bibitem{LOD.survey}
Meusel, R., Bizer, C., Paulheim, H.: A web-scale study of the adoption and
  evolution of the schema.org vocabulary over time. In: Intl Conf on Web
  Intelligence, Mining and Semantics. ACM, New York, NY, USA (2015).
  \doi{10.1145/2797115.2797124}

\bibitem{miller}
Nargesian, F., Zhu, E., Pu, K.Q., Miller, R.J.: Table union search on open
  data. VLDB Journal  \textbf{11}(7) (Mar 2018). \doi{10.14778/3192965.3192973}

\bibitem{sdata}
Nature scientific data. \url{https://www.nature.com/sdata} (2018)

\bibitem{www2019}
Noy, N., Burgess, M., Brickley, D.: {Google Dataset Search}: Building a search
  engine for datasets in an open web ecosystem. In: The Web Conference. p.
  1365–1375. ACM (2019). \doi{10.1145/3308558.3313685}

\bibitem{kg.cacm}
Noy, N., Gao, Y., Jain, A., Narayanan, A., Patterson, A., Taylor, J.:
  Industry-scale knowledge graphs: Lessons and challenges. Commun. ACM
  \textbf{62}(8),  36–43 (Jul 2019). \doi{10.1145/3331166}

\bibitem{RDF}
{RDF 1.1 Concepts and Abstract Syntax}.
  \url{https://www.w3.org/TR/rdf11-concepts/}

\bibitem{datacite}
Rueda, L., Fenner, M., Cruse, P.: Datacite: Lessons learned on persistent
  identifiers for research data. {IJDC}  \textbf{11}(2),  39--47 (2016).
  \doi{10.2218/ijdc.v11i2.421}

\bibitem{dats}
Sansone, S.A., Gonzalez-Beltran, A., Rocca-Serra, P., Alter, G., Grethe, J.S.,
  Xu, H., Fore, I.M., Lyle, J., Gururaj, A.E., Chen, X., et~al.: {DATS}, the
  data tag suite to enable discoverability of datasets. Scientific data
  \textbf{4},  170059 (2017)

\bibitem{Schmachtenberg}
Schmachtenberg, M., Bizer, C., Paulheim, H.: Adoption of the linked data best
  practices in different topical domains. In: Intl Semantic Web Conf (ISWC).
  pp. 245--260. Springer (2014)

\bibitem{Stall.Nature}
Stall, S., Yarmey, L., Cutcher-Gershenfeld, J., Hanson, B., Lehnert, K., Nosek,
  B., Parsons, M., Robinson, E., Wyborn, L.: Make scientific data {FAIR} (2019)

\bibitem{vrandecic.akts}
Vrande{\v{c}}i{\'c}, D.: Describing datasets in {Wikidata}. In: {Advanced
  Knowledge Technologies for Science in a FAIR World, IEEE eScience Conference}
  (2019)

\bibitem{wang2017}
Wang, J., Aryani, A., Wyborn, L., Evans, B.: Providing research graph data in
  {JSON-LD Using Schema.org}. In: 26th Intl Conf on World Wide Web Companion.
  pp. 1213--1218 (2017). \doi{10.1145/3041021.3053052}

\bibitem{FAIR}
Wilkinson, M.D., Dumontier, M., Aalbersberg, I.J., et~al.: The {FAIR} guiding
  principles for scientific data management and stewardship. Scientific data
  \textbf{3} (2016)

\bibitem{compact.identifiers}
Wimalaratne, S.M., Juty, N., Kunze, J., Jan{\'e}e, G., et~al.: Uniform
  resolution of compact identifiers for biomedical data. Scientific data
  \textbf{5},  180029 (2018)

\end{thebibliography}

\end{document}